\documentclass[cameraready]{Interspeech}

\title{Ethical and Technical Limits of Deepfake Speech Datasets}

\author[orcid=0009-0000-5722-0571]{Vojtěch}{Staněk}
\author{Eva}{Trnovská}
\author[orcid=0000-0002-9009-2193]{Kamil}{Malinka}
\author[orcid=0000-0002-4717-1910]{Anton}{Firc}

\address{
    Security@FIT, Brno University of Technology, Czech Republic
}

\email{istanek@fit.vut.cz, xtrnov01@stud.fit.vut.cz, malinka@fit.vut.cz, ifirc@fit.vut.cz}

\keywords{audio deepfakes, spoofing countermeasures, dataset audit, fairness evaluation, cross-dataset generalization}

\usepackage{comment}
\usepackage{enumitem}

\begin{document}

\maketitle

\begin{abstract}
    Claims about the robustness and fairness of deepfake speech detectors are only as credible as the datasets used to train and evaluate those systems. We present a dataset-level audit of the deepfake speech landscape. We compile and analyze 39 deepfake speech datasets, examining key attributes including accessibility, documentation, demographic and language coverage, dataset scale, and the underlying bona fide speech sources. Our audit reveals two important takeaways. Firstly, fairness assessment is largely infeasible because most datasets lack demographic metadata, and only a few contain gender or language labels. This prevents any meaningful subgroup analysis and leaves other demographic attributes unaddressed. Secondly, we identify substantial overlap in underlying bona fide source corpora across datasets, which can undermine cross-dataset evaluation and lead to overstated generalization claims. 
\end{abstract}

\section{Introduction}



The misuse of synthetic speech poses a growing challenge for speaker verification, digital forensics, and media authentication~\cite{Firc2025, europol_reality, Malinka202466}. Employing a deepfake speech detector is a core defense mechanism~\cite{Firc2024, evolutionary_fusion}, but moving toward real-world usage shifts focus from accuracy alone to robustness, fairness, and transparency. In high-stakes settings, detectors should be non-discriminative across demographic groups, as biased performance can disproportionately affect marginalized communities and raise human rights concerns.

Furthermore, emerging regulations, such as the EU AI Act~\cite{EU_AI_Act_2024}, increasingly emphasize dataset documentation, traceability, and bias monitoring. However, current deepfake speech datasets were primarily created for benchmarking detection performance~\cite{asvspoof5, voicewukong}. This creates a mismatch between practical necessities and what existing benchmarks actually support.

Fairness and bias assessment require demographic metadata, yet most deepfake speech datasets provide only limited or no such annotations. Without such information, disparities across speakers, accents, or languages cannot be measured or mitigated. This makes bias assessment infeasible beyond a small subset of resources with limited coverage~\cite{MLAAD, scdf}.

Moreover, generalization is frequently assessed by training on one benchmark and testing on another. However, as shown in \autoref{fig:source-df}, many benchmarks reuse the same bona fide source corpora, meaning that \textit{cross-dataset} evaluation does not necessarily imply \textit{out-of-domain} evaluation. This shared source can cause data leakage: detectors may exploit corpus-specific artifacts rather than deepfake cues, thereby overestimating robustness and out-of-domain generalization.

Therefore, this paper provides an audit of deepfake speech datasets\footnote{Interactive browser of the dataset table and provenance map: \url{https://security-fit.github.io/deepfake_speech_datasets_app/}}. We compile and analyze 39 datasets reported in peer-reviewed publications (\autoref{tab:datasets}) and map overlaps across the underlying bona fide source corpora (\autoref{fig:source-df}) to show current limitations in addressing bias.
We hope this work sparks a discussion on audit-ready dataset practices that enable fair, non-discriminatory, and interpretable deepfake speech detection evaluation, helping prevent demographic biases and enabling safer deployment in high-stakes settings.

\textbf{Contributions.} Our contributions are twofold:
\begin{itemize}[topsep=0pt, itemsep=0pt]
    \item We provide a structured audit of 39 deepfake speech datasets, analyzing accessibility, language, demographic metadata, synthesis documentation, and dataset scale.
        \item We map bona fide source overlap, revealing common dependence on a small subset of speech corpora and the resulting risks for cross-dataset generalization claims.
\end{itemize}


\section{Background}

Current work on deepfake speech datasets focuses on model benchmarking~\cite{asvspoof5,voicewukong} or high-level dataset overviews~\cite{firc2023deepfakes}, typically reporting basic properties such as size, year, and sometimes a short description, with limited attention to dataset auditability and representativeness (e.g., demographic metadata or data sources). In parallel, the broader ML community has proposed structured dataset documentation~\cite{siddik2025datasheetshealthcareaiframework} to improve transparency and enable bias auditing; however, recent studies reveal that dataset documentation remains inconsistent and omits key information such as data provenance or demographic metadata~\cite{liang2024whatsdocumentedaisystematic,oreamuno-datasets,yang2024navigatingdatasetdocumentationsai}. 
Finally, fairness disparities in deepfake detection have been documented in the facial domain~\cite{gbdf-dataset,Liu2024Thinking,Hazirbas2022Towards}. While some insight into the fairness of deepfake speech detectors is provided by initial studies on gender~\cite{Yadav2024FairSSD} or language~\cite{MLAAD}, a dataset-level audit of deepfake speech resources has received limited attention, especially regarding representativeness, fairness, and dataset biases~\cite{scdf}. This paper addresses this gap by auditing 39 deepfake speech datasets and mapping the underlying source-corpora overlap to highlight how missing metadata and source overlap can critically undermine claims about generalization and fairness of deepfake speech detectors.


\section{Audit of Existing Deepfake Speech Datasets}
\label{sec:analysis}


\begin{table*}[!ht]
\centering
\caption{Overview of important properties of deepfake speech datasets (\textit{$\star$} denotes uncertainty arising from a vague description of respective datasets).}
\begin{tabular}{@{}llllrrrl@{}}
\textbf{Dataset} & \textbf{Year} & \textbf{Accessibility} & \textbf{DF tools} & \textbf{Lang.} & \textbf{Utterances} & \textbf{Speakers (M+F)} & \textbf{License}\\
\midrule
VCC 2016 \cite{VCC2016} & 2016 & public & 18 & 1 & 25 110 & 10 (5+5) & CC BY 4.0 \\
VCC 2018 \cite{VCC2018} & 2018 & public & 23 & 1 & 41 148 & 12 (6+6) & CC BY 4.0 \\
ASVspoof 2019 LA \cite{ASVspoof2019} & 2019 & public & 17 & 1 & 121 461 & 107 (46+61) & ODC-By \\
Fake or Real \cite{FoR} & 2019 & public & 7 & 1 & 198 000 & N/A & GPLv3 \\
MC-TTS \cite{DeepSonar} & 2020 & restricted & 1 & 1 & 12 026 & N/A & N/A \\
Sprocket-VC \cite{DeepSonar} & 2020 & restricted & 1 & 1 & 6 588 & N/A & N/A \\
SynSpeechDDB \cite{SynSpeechDDB} & 2020 & restricted & 16 & 2 & 142 290 & N/A & N/A \\
VCC 2020 \cite{VCC2020} & 2020 & public & 3 & 4 & 30 280 & 14 (7+7) & DbCL 1.0 \\
ASVspoof 2021 DF \cite{ASVspoof2021} & 2021 & public & 100$\star$ & 1 & 611 829 & 93 (43+50) & ODC-By \\
ASVspoof 2021 LA \cite{ASVspoof2021} & 2021 & public & 13  & 1 & 164 640 & 67 (30+37) & ODC-By \\
FMFCC-A \cite{FMFCC-A} & 2021 & restricted & 13 & 1 & 50 000 & 131 (58+73) & N/A \\
Half-Truth (HAD) \cite{Half-Truth} & 2021 & public & 1 & 1 & 160 836 & 218 (43+175) & CC BY 4.0 \\
WaveFake \cite{Wavefake} & 2021 & public & 7 & 2 & 122 985 & 2 (0+2) & CC BY-SA 4.0 \\
AD for SFR \cite{SFR} & 2022 & public & 5 & 1 & 181 764 & 585$\star$ (N/A) & CC BY 4.0 \\
ADD challenge 1 \cite{ADD1} & 2022 & public & N/A & 1 & 493 123 & 80 (40+40) & CC BY-NC-ND 4.0 \\
CFAD \cite{CFAD} & 2022 & public & 11 & 1 & 347 514 & 1212 (N/A) & CC BY 4.0 \\
F\&M \cite{FM} & 2022 & public & 1 & 2 & 3 522 & 160 (N/A) & N/A \\
In-the-wild \cite{In-the-wild} & 2022 & public & N/A & 1 & 31 752 & 58 (N/A) & Apache 2.0 \\
ADD challenge 2 \cite{ADD2} & 2023 & public & N/A & 1 & 517 068 & 1030$\star$ (N/A) & CC BY-NC-ND 4.0 \\
DECRO \cite{DECRO} & 2023 & public & 10 & 2 & 118 382 & 2839 (N/A) & CC BY 4.0 \\
LibriTTS-DF \cite{SWAN-DF} & 2023 & restricted & 6 & 1 & N/A & N/A & N/A \\
PartialSpoof v1.2 \cite{PartialSpoof} & 2023 & public & 9$\star$ & 1 & 121 461 & 107$\star$ (46+61) & CC BY 4.0 \\
TIMIT-TTS \cite{TIMIT-TTS} & 2023 & public & 12 & 1 & 79 120 & 69 (38+31) & CC BY 4.0 \\
Voc.v2,v3,v4 \cite{Vocv} & 2023 & public & 4 & 1 & 10,320$\star$ & 20 (8+12) & ODC-By \\
MLAAD (v9) \cite{MLAAD} & 2024 & public & 140 & 51 & 298 000 & N/A & CC BY-NC 4.0 \\
CodecFake \cite{CodecFake} & 2024 & public & 15 & 1 & 90 797 & 109 (47+62) & CC BY 4.0 \\ 
Codecfake \cite{Yi2024Codecfake_big} & 2024 & public & 7 & 2 & 1 058 166 & 328 (N/A) & CC BY-NC-ND 4.0 \\
CVoiceFake \cite{CVoiceFake} & 2024 & public & 6 & 5 & 1 254 893$\star$ & N/A & CC BY 4.0 \\
Diffuse or Confuse \cite{Firc2024Diffuse} & 2024 & public & 12 & 1 & 183 400 & 1 (0+1) & CC BY 4.0 \\
MLADDC \cite{shah2024mladdc} & 2024 & public & 2$\star$ & 20 & 400 000 & N/A & CC BY 4.0 \\
VoiceWukong \cite{voicewukong} & 2024 & restricted & 34 & 2 & 826 800 & N/A & CC BY-NC 4.0 \\
DFADD \cite{du2024dfadd} & 2024 & public & 5 & 1 & 207 995 & 109 (47+62) & MIT \\
ASVspoof 5 \cite{asvspoof5} & 2024 & public & 32 & 1 & 1 004 081 & 1922 (964+958) & ODC-By \\
SCDF \cite{scdf} & 2025 & public & 4 & 5 & 237 250 & 50 (25+25) & CC0 1.0 \\
AI4T \cite{ai4t} & 2025 & public & N/A & 8 & 4 798 & N/A & N/A \\
SynHate \cite{synhate} & 2025 & public\textit{$\star$} & N/A & 37 & 134 797 & N/A & N/A \\
STOPA \cite{stopa} & 2025 & public & 13 & 1 & 699 000 & 107 (46+61) & CC BY 4.0 \\
PartialEdit \cite{partialedit} & 2025 & public\textit{$\star$} & 4 & 1 & 43 358 & 107\textit{$\star$} (46+61) &  CC BY 4.0\\
SpeechFake \cite{speechfake} & 2025 & public & 30 & 46 & 3 338 508\textit{$\star$} & 720 (374+346)\textit{$\star$} & Apache 2.0 \\
\midrule
\end{tabular}
\vspace{-1.47em}
\label{tab:datasets}
\end{table*}

This section audits existing deepfake speech datasets to evaluate whether available resources are suitable for fairness evaluation and cross-dataset generalization testing. 
Our contribution is a dataset-level audit that identifies gaps in demographic balance and in the diversity of synthesis tools, revealing potential evaluation biases in reported results.
We compiled a list of 39 datasets by analyzing prior surveys and commonly referenced benchmarks, then manually extended it using keyword-based searches to identify additional datasets reported in peer-reviewed publications.

We selected datasets designed for deepfake speech detection that include bona fide and synthesized speech (TTS and/or VC) and provide public documentation for attribute extraction. We excluded datasets that either do not directly contain deepfake speech, pre-date modern synthesizers (before 2018)\footnote{VCC 2016 and VCC 2018 are included because they are directly reused by more modern datasets.}, or have very low sample counts (below 1,000), as these are unlikely to support meaningful detector training or evaluation.
We analyze key dataset parameters to assess their suitability for robust and fair evaluation from technical, ethical, and practical perspectives. Attributes were cross-checked against the connected publication and, when available, the official repository/project page. The results are presented in \autoref{tab:datasets}.

\textit{1) Synthesizers:} A diverse range of synthesizers is arguably the most critical factor for effectively training detectors to address real-world scenarios, particularly since utterances generated by novel methods may evade detection~\cite{prudky2023, comprehensivemultiparametric}. The examined datasets list more than 100 distinct tools and architectures (both open-source and commercial), indicating significant diversity. Recordings are generated by both VC and TTS tools. 
However, not all datasets specify the tools used, which significantly hinders understanding of generalization ability during cross-evaluation, especially when the training and evaluation datasets use recordings produced by similar technologies.

\begin{figure*}[!ht]
    \centering
    \includegraphics[width=0.981\textwidth]{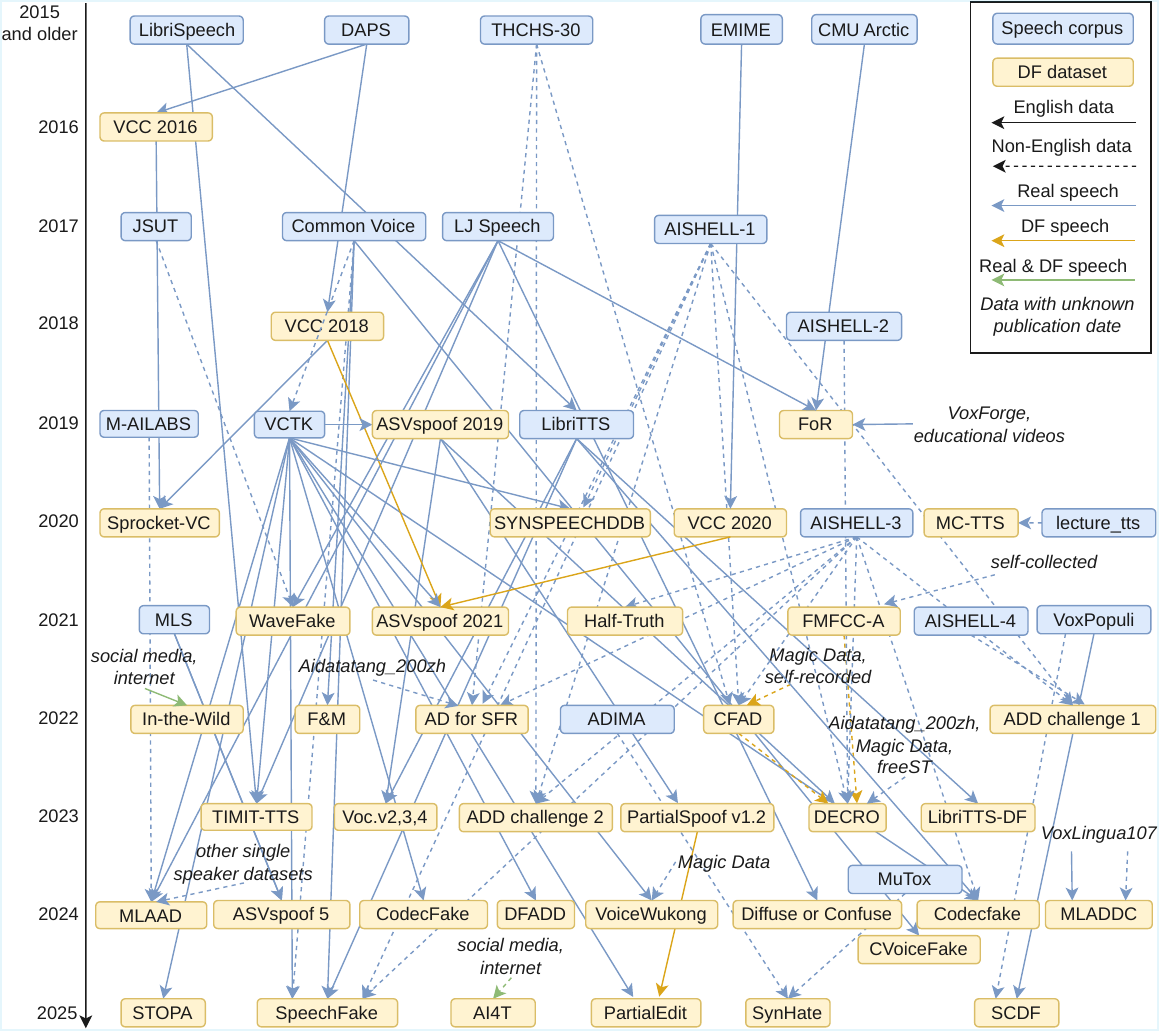}
    \caption{Audio datasets containing deepfake speech with the corpus they were derived from -- utterances were taken directly from the corpus, or the corpus was used to train synthesis tools if mentioned in the paper. The time axis (Y) shows the year of publication. Best viewed in color.}
    \label{fig:source-df}
    \vspace{-1.5em}
\end{figure*}

\textit{2) (Bona fide) speech sources:} The origin of the bona fide speech plays a crucial role in assessing the generalization ability of deepfake speech detectors. Deepfake datasets draw from a wide range of sources, typically established speech corpora and seldom from social media; purely self-recorded data is rare. Importantly, many datasets are derived from overlapping sets of bona fide speech sources. Notably, LibriVox~\cite{LibriVox} serves as a source for multiple corpora, including LibriTTS~\cite{libritts}, LJSpeech~\cite{ljspeech17}, LibriSpeech~\cite{LibriSpeech}, and Multilingual LibriSpeech (MLS)~\cite{MLS}, which are further used as a foundation for many deepfake speech datasets. Most often, datasets are derived from the LJSpeech~\cite{ljspeech17}, AISHELL~\cite{aishell_2017}, and VCTK~\cite{VCTK} corpora, as presented below in \autoref{fig:source-df}.

This reliance on shared source data discredits the validity of cross-dataset evaluations. When multiple datasets are built on the same underlying bona fide corpora, cross-dataset performance may be inflated if models exploit source-corpus characteristics. Consequently, a detector's true robustness and its ability to generalize to unseen deepfakes might be overstated. Understanding and accounting for this source overlap is essential for accurate and truthful evaluation.

While we map the reported underlying bona fide source corpora across datasets, precise quantification of overlap is generally not possible from public documentation. Many datasets do not disclose the necessary information about how the source corpus was used, which speaker IDs were included, or the train/dev/test partitioning, and some repackage audio with additional filtering, re-segmentation, or other preprocessing steps. Therefore, our map should not be taken as an authoritative ground-truth measure of overlap, but rather as an indicator of potential bias in cross-dataset training and evaluation.

\textit{3) Demographics:} Demographic metadata is essential for analyzing and mitigating bias in deepfake speech detection. Without adequate demographic data, ensuring the effectiveness of detection solutions across diverse groups becomes unattainable. The existing datasets currently do not sufficiently incorporate this essential information. Only some datasets provide protocols or metadata that enable controlled splits~\cite{MLAAD,scdf}; in other cases, limited attributes can be inferred indirectly (e.g., gender from the folder structure~\cite{VCC2018}). Only a small subset provides an explicitly balanced gender composition~\cite{scdf,VCC2020,speechfake}. Beyond gender and language, demographic information is largely absent. Attributes such as accent, age, ethnicity, or disability factors are not reported or considered, limiting the ability to assess representativeness and bias. The absence of demographic metadata is the main limiting factor for fairness evaluation.

\textit{4) Language:} Most deepfake speech datasets are monolingual, mostly English and Chinese. Only a handful are bilingual~\cite{voicewukong,Wavefake}. Multilingual datasets have recently emerged, including MLAAD~\cite{MLAAD}, AI4T~\cite{ai4t}, SynHate~\cite{synhate}, SCDF~\cite{scdf} and SpeechFake~\cite{speechfake}. These datasets are also trying to provide balanced language subsets. Prior results indicate that detector performance drops on previously unseen languages~\cite{DECRO}, underscoring the need for multilingual, well-documented datasets to evaluate linguistic bias. Moreover, datasets in languages other than English and Chinese are still lacking.

\textit{5) Size:} Dataset scale affects the feasibility of training modern detectors. Small datasets are unlikely to be suitable for DNN-based training; prior work suggests that fewer than 80,000 samples can be insufficient in some settings~\cite{Firc2024}. However, the sample count alone is not informative without adequate diversity in speakers and synthesis methods.

\textit{6) Publication year:} Because speech synthesis quality evolves rapidly, staying up to date with this trend is crucial. Older datasets may under-represent current attacks, leading to evaluations that do not reflect state-of-the-art deepfakes.

\textit{7) Availability:} Access and licensing remain practical barriers: 15\% (6/39) of datasets are restricted, and some are described as public but are not obtainable via the provided links~\cite{synhate} or are only partially released~\cite{partialedit}. Even when accessible, restrictive or unclear licenses can still limit reproducibility and practical deployment.




\section{Findings \& Discussion}

Our dataset audit uncovered major shortcomings of the currently available deepfake speech datasets. Most issues originate from missing metadata, undisclosed synthesis details, insufficient gender and language coverage, outdated synthesis tools, and licensing or access constraints.

Only 19/39 (49\%) datasets report speaker counts and metadata labels for both female and male speakers, substantially limiting the evaluation of gender bias. More broadly, fairness assessment in deepfake speech detection is often dataset-limited: without reliable demographic labels (and ideally balanced representation), it is not possible to quantify disparities between demographic subgroups or to support fairness claims.

In terms of language coverage, datasets contain mostly English or Chinese speech. The majority of datasets (25/39; 64\%) are monolingual; a few are bilingual (6/39; 15\%); and only recently have multilingual datasets begun to emerge (8/39; 21\%), indicating that linguistic diversity remains an exception rather than the standard. This is particularly problematic given evidence that detector performance degrades on unseen languages~\cite{MLAAD,DECRO}, making linguistic bias difficult to measure without multilingual, well-documented benchmarks.

Beyond metadata availability, we reveal that evaluation validity is threatened by a shared source of bona fide speech. As summarized in \autoref{fig:source-df}, many deepfake speech datasets are constructed from an overlapping set of speech corpora (mainly LJSpeech, VCTK, AISHELL, and LibriVox-derived resources). This overlap discredits cross-dataset evaluations, as models may exploit specific corpus artifacts that can artificially inflate reported generalization results. As a direct consequence, robustness to truly unseen speech (both deepfake and bona fide) may be overstated unless this overlap is explicitly addressed.

Regarding synthesis documentation, 9/39 (23\%) datasets either contain only a single synthesizer tool or do not disclose tool information, which complicates reproducibility and cross-dataset interpretation. Modern attacks evolve rapidly, and targeting older synthesizers makes the detector vulnerable to realistic deepfakes of the current time. Additionally, incomplete disclosure of dataset synthesizers and attacks makes it difficult to analyze failure cases, which could help detector developers mitigate errors and create a more robust detection system.

Finally, public availability and licensing remain practical barriers: 6/39 (15\%) datasets are not publicly available at all. Some datasets (8/39; 21\%) lack a license, creating significant legal uncertainty about whether they can be used for research or commerce. From the public datasets, 4/33 (12\%) are released under a license that forbids commercial use, while some specific licenses (GPL, CC BY-SA) may require derived resources to be released under the same open license, which might be a significant problem in the commercial field. Moreover, legal aspects may vary across countries. Luckily, the majority of public datasets (26/33; 79\%) specify a permissive license which allows utilization in both research and commerce. 


Together, our findings suggest that progress in deepfake speech detection is increasingly constrained not only by model design but also by the representativeness of datasets. There are two key takeaways from the audit. Firstly, non-discriminatory deepfake detection is hindered primarily by the lack of demographic metadata and by non-representative datasets that do not support fairness evaluation beyond gender or, less often, language. Other demographic attributes are typically absent. Secondly, contrary to common evaluation practice, perceived cross-dataset generalization might be biased by shared bona fide source corpora. Shared provenance can introduce data leakage, undermining the claimed \textit{out-of-domain} evaluation and leading to overstated robustness.

Importantly, none of the audited datasets satisfies all the discussed properties for unbiased deepfake detection. A common workaround is to combine multiple datasets; however, this introduces additional biases into the evaluation and may still not cover the diversity required for robust deployment. As a result, reported fairness claims or cross-dataset generalization should be interpreted cautiously unless the mentioned biases are explicitly addressed.

\section{Conclusion}

This paper presented a dataset-level audit of 39 deepfake speech datasets and mapped overlap in underlying bona fide source corpora. Our analysis shows that missing demographic and language metadata makes fairness and bias evaluation infeasible with the resources currently available. Additionally, we identified substantial overlap between source corpora, which can undermine cross-dataset evaluation and overstate generalization. Finally, incomplete synthesis documentation, restricted access, or unclear licensing limit reproducibility and practical adoption. 

We therefore recommend that dataset releases and connected benchmark papers report demographic and language metadata, disclose detailed synthesis pipelines, document the source of bona fide data, provide clear licensing, and ensure reliable access to released datasets. These practices are crucial for enabling reproducible evaluation and for developing fair and reliable deepfake speech detectors usable in practical scenarios. Finally, we release an interactive browser to support community inspection of provenance overlap and dataset properties, reducing the need for extensive manual research across papers and repositories: {\footnotesize \url{https://security-fit.github.io/deepfake_speech_datasets_app/}}.

\section{Acknowledgments}

This work was supported by the Brno University of Technology internal project FIT-S-26-9011, and the Ministry of Education, Youth and Sports of the Czech Republic through the e-INFRA CZ project (ID: 90254).

\section{Generative AI Use Disclosure}

During the preparation of this work, the authors used Generative AI Models (specifically Google Gemini, ChatGPT, and Grammarly) for language editing and text refinement. The authors reviewed and edited the output as needed and take full responsibility for the publication's content.

\bibliographystyle{IEEEtran}
\bibliography{mybib}

\end{document}